\documentstyle[multicol,aps,prl,epsf]{revtex}

\begin{document}
\pagestyle{empty}

\begin{multicols}{2}
\narrowtext
\parskip=0cm

\noindent
{
\bf 
Comment on ``Evidence for the Drop\-let/Scaling Picture of
Spin Glasses''}
\smallskip

In a recent letter Moore et al. \cite{MOBODR} claim to exhibit
evidence for a non-mean-field behavior of the $3d$ Ising spin
glass. We show here that their claim is insubstantial, and by
analyzing in detail the behavior of the Migdal-Kadanoff approximation
(MKA) as compared to the behavior of the Edwards-Anderson (EA) spin
glass we find further evidence of a mean-field like behavior of the
$3d$ spin glass.

The main point of \cite{MOBODR} does not concern the validity of the
MKA in describing spin glasses, since it is well known, after the work
of \cite{GARDNER}, that already at the mean field level the MKA
describes a trivial droplet structure, completely missing the
structure of the phase space of the model in any dimension.

Reference \cite{MOBODR} shows instead that the probability
distribution of the order parameter $P_{MK}(q)$ computed in the MKA at
$T=0.7$, close to the temperature where most of the numerical
simulations have been run \cite{NUMREV}, has a spurious small $q$
``plateau'', very similar to the non-trivial $P(q)$ one finds
numerically for the EA model.  In these conditions, for values of the
lattice size comparable to the ones used in numerical simulations,
$L\le 16$, the small $q$ region of $P_{MK}(q)$ does not seem to depend
on $L$, even if one knows that eventually, for very large values of
$L$, it will have to become trivial. The authors of \cite{MOBODR}
explain this coincidence as a hint of the fact that asymptotically the
EA model will also behave as a droplet model.

Here we show that this similarity in the behavior of the MKA and the
true EA $3d$ spin glass does not concern observables that are crucial
for determining replica symmetry breaking (RSB). We look at the {\em
link overlap} (on a system of linear size $L$ and volume $V=L^3$)
$q^{(L)}\equiv 1/(3V) \sum' \langle \sigma_i \sigma_{i+\hat{\mu}}
\tau_i \tau_{i+\hat{\mu}} \rangle $, where the sum runs over
first-neighbor site pairs.  $q^{(L)}$ is more sensitive than the usual
overlap $q$ to the difference between a droplet and a mean field like
behavior.  The link overlap is, as discussed for example in
\cite{NUMREV}, of crucial importance, since a non-trivial $P(q)$ could
be simply due to the presence of interfaces, while a non-trivial
$P(q^{(L)})$ is a non-ambiguous signature of RSB.

We show that one can see a clear difference, already at $T=0.7$ on
medium-size lattices, among the MKA and the EA model.  So, not only
our observation makes the point of \cite{MOBODR} obsolete, but it also
shows that simulations on reasonable-sized lattices are useful, when
studying either disordered systems or normal statistical mechanical
models (from the point of view of the advocates of \cite{MOBODR} in
the case of disordered systems only simulations on systems of huge
size could make the true nature of the system manifest).

We have analyzed the MKA of the $3d$ spin glass (averaging over $1000$
disorder samples), and the $3d$ EA model by numerical simulations
(using a tempering algorithm \cite{NUMREV} and an annealing scheme,
checking convergence and averaging over $64$ or more samples). In all
cases we have considered binary couplings and a Hamiltonian
$H_\epsilon[\sigma,\tau] = H_0[\sigma] + H_0[\tau] - \epsilon
\sum' \sigma_i \sigma_{i+\hat{\mu}} \tau_i \tau_{i+\hat{\mu}}$, 
where $H_0$ is the usual EA $3d$ Hamiltonian.

\begin{figure}
\centerline{
\epsfysize=0.7\columnwidth{\epsfbox{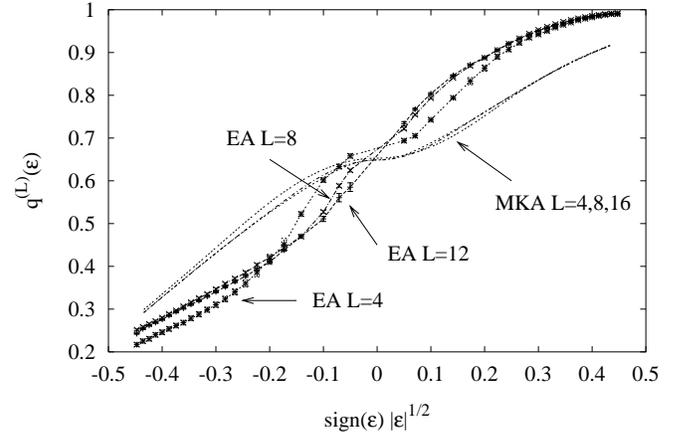}}}
\narrowtext{
\caption{ 
$q^{(L)}$ in the MKA (lines without points) 
and from simulations of the $3d$ EA spin glass.
\protect\label{fig1}
}  
}
\end{figure}

In fig. (\ref{fig1}) we show our results for $q^{(L)}(\epsilon)$
versus $\epsilon^{\frac12}$.  The MKA gives a smooth behavior: for
small $\epsilon$, $q^{(L)}(\epsilon)$ behaves like
$\epsilon^{\lambda}$, with $\lambda\simeq 1$.  Finite size effects
look very small for these sizes (from $4$ to $16$).  The EA model
behaves in a completely different way.  Here finite size effects are
large, and the behavior for small $\epsilon$ becomes more singular for
larger sizes.  The $L=4$ lattice is reminiscent of the MKA behavior,
but already at $L=8$ the difference is clear.  From our data we are
not able to definitely establish the existence of a discontinuity, but
the numerical evidence is strongly suggestive of that.  The data are
suggestive of the building up of a discontinuity as $L\to\infty$, i.e.
$q=q_{+}+A_{+}\epsilon^{\lambda}$ for $\epsilon>0$ and
$q=q_{-}+A_{-}|\epsilon|^{\lambda}$ for $\epsilon<0$, with $q_{+}\ne
q_{-}$ and an exponent $\lambda$ close to $\frac12$: a continuous
behavior (i.e.  $q_{+}= q_{-}$) cannot be excluded from these data,
but in this case we find an upper limit $\lambda < 0.25$, totally
different from the behavior of MKA, $\lambda \simeq 1$.  This is what
is needed to show that when looking at observables that are very
sensitive to RSB the difference among the trivial behavior of the MKA
and true spin glasses is already clear at $T\simeq 0.6 T_c$ on
lattices of size $L\simeq 16$, as opposed to the claims of
\cite{MOBODR}.

\noindent 
E. Marinari, G. Parisi, J. J. Ruiz-Lorenzo and F. Zuliani


%

\end{multicols}
\end{document}